\documentclass[11pt, a4paper]{article}
\pdfoutput=1 

\usepackage{jheppub} 

\usepackage[T1]{fontenc} 
\newcommand{\ex}[1]{\langle x^{#1} \rangle}
\newcommand{\er}[1]{\langle r^{#1} \rangle}
\newcommand{\OO}{\mathcal{O}}

\title{ Bootstrapping Simple QM Systems}

\author{David Berenstein $^\dagger$}
\author{and George Hulsey $^\ddagger$}
\affiliation{Dept. of Physics, University of California, Santa Barbara\\Santa Barbara, CA 93106}

\emailAdd{$^\dagger$ dberens@physics.ucsb.edu}
\emailAdd{$^\ddagger$ hulsey@physics.ucsb.edu}

\abstract{We test the bootstrap approach for determining the spectrum of one dimensional Hamiltonians, following the recent approach of Han, Hartnoll, and Kruthoff. We focus on comparing the bootstrap method data to known analytical predictions for the hydrogen atom and the harmonic oscillator. We resolve many energy levels for each, and more levels are resolved as the size of the matrices used to solve the problem increases. Using the bootstrap approach we find the spectrum of the Coulomb and harmonic potentials converge exponentially fast. }

\begin{document} 
\maketitle
\flushbottom

\section{Introduction}
\label{sec:intro}

Most quantum mechanical systems, even in one dimension, are not soluble by analytical means. Finding novel methods for solving these systems numerically is always useful. Recently, a booststrap method for solving quantum mechanical systems has been introduced in \cite{hart}. Our goal in this paper is to explore how well the bootstrap works in some problems that have an analytical solution. This is then a test of the effectiveness of the method to solve other problems.

The basic idea of the quantum mechanical bootstrap is very much the same as familiar bootstrap programs in CFT, which have led to high precision numerical solutions of the Ising model in 3 dimensions \cite{El-Showk:2012cjh}. More recently this proposal has been used to investigate matrix models which have both analytical and numerical solutions that can be compared \cite{Kazakov:2021lel}.

The method works as follows: initial guesses are made for some parameters (data) of the model, and from these guesses a list of predictions based on the dynamics is made. We apply consistency checks to these predictions recursively and reject initial guesses which fail the checks at a given step. These follow from two basic identities. First, on an eigenstate of the Hamiltonian it is true that certain expectation values vanish identically
\begin{equation}
    \langle [H, {\cal O}]\rangle=0. \label{eq:comm1}
\end{equation}
It is also true that, in states of energy $E$, we have
\begin{equation}
    \langle H {\cal O}\rangle= E \langle {\cal O}\rangle \label{eq:en1}
\end{equation}
Secondly, one uses positivity constraints like
\begin{equation}
 \langle {\cal O}^\dagger {\cal O}\rangle\geq 0
\end{equation}
which holds for any operator ${\cal O}$. Importantly, the expectation values of the squares of certain operators appear  from applying (\ref{eq:comm1}), (\ref{eq:en1}) recursively, so one has a consistent set of equations to solve. This is done by guessing $E$ and perhaps some other data.

In this way one hopes that the solutions that pass all the tests will eventually converge to actual solutions of the theory, and from this we can learn nontrivial facts about the theory in question. In practice, with finitely many constraints, one hopes that the allowed region becomes small enough that one can learn a lot about the theory. This is not guaranteed. In our examples we will see this convergence in most cases, but when it happens is not clear \textit{a priori}. We also see that in principle, with enough computational resources, we can  recover all the (bound state) energy levels of these systems with high precision.

Some of the original ideas can be traced to work in the Hamiltonian formulation of large $N$ gauge theories, in particular constructing an effective field theory of gauge invariant collective coordinates 
\cite{Jevicki:1982jj,Jevicki:1983wu,Rodrigues:1985aq}. In this setup, together with factorization, the problem reduces to finding the minima of an effective potential and checking that the kinetic term is positive definite. 

This approach was applied to matrix integrals recently by Lin \cite{lin} and applied to quantum mechanics and matrix QM by Hartnoll et al \cite{hart}. In the interest of testing the method's performance, we apply it to quantum mechanics problems with well-known analytical predictions. More recently, in \cite{dirac}, the authors employ similar techniques to bootstrap ensembles of Dirac operators. Previous attempts to use similar positivity methods for gauge theories can be found in \cite{Anderson:2016rcw,Anderson:2018xuq}.

In this paper we focus on the simplest possible cases of the quantum mechanical bootstrap: the Coulomb potential/hydrogen Hamiltonian and the simple harmonic oscillator. These are the simplest models to bootstrap in the sense that the trial parameter space is one-dimensional. They also admit analytical solutions, which allows us to test the efficacy and accuracy of the method. We will characterize the accuracy, precision, and convergence of the bootstrap method for these one-dimensional quantum mechanical problems; i.e. finding eigenvalues for the time-independent Schr{\"o}dinger equation.

\section{Bootstrapping quantum mechanics}
For our focus on one-dimensional quantum mechanics, we start with some Hamiltonian 
\begin{equation}\label{eq:genH}
    H = \frac{p^2}{2M} + V(x)
\end{equation}
To each energy eigenstate of $H$ is associated an energy $E$ and a sequence of moments $\{\ex{n}\}_0^\infty$. Knowing the energy and all the moments is equivalent to knowing the PDF associated to the wavefunction of some given eigenstate. The goal of the quantum mechanical bootstrap is to approximately identify the energies and moments $\ex{n}$ corresponding to the eigenstates of $H$. 

\subsection{Moment recursion}
We start by applying the method of \cite{hart} to obtain a recursion relation for the positional moments. For any Hamiltonian $H$, in any energy eigenstate, we have the following identities of expectation values for an operator $\OO$:
\begin{equation}
	\langle [H,\OO] \rangle = 0 \qquad \langle H\OO \rangle = E\langle \OO \rangle
\end{equation}
Throughout, angle brackets denote an expectation value in an arbitrary energy eigenstate of $H$. These identities allow us to relate different moments of $x,p$. For example, take $\OO= x^s$ and let $H$ be the generic Hamiltonian (\ref{eq:genH}) with $M = 1$. Using the first identity above gives a relation
\begin{equation*}
	\langle [H,x^s] \rangle =0\quad \implies \quad s\left\langle x^{s-1} p\right\rangle=\frac{i}{2} s(s-1)\left\langle x^{s-2}\right\rangle
\end{equation*}
This does not depend on the potential $V(x)$. If we similarly take $\OO = x^mp$, we get such dependence:
\begin{equation*}
	0=m\langle x^{m-1} p^{2}\rangle+\frac{1}{4} m(m-1)(m-2)\langle x^{m-3}\rangle-\langle x^{m} V^{\prime}(x)\rangle
\end{equation*}
Finally we can involve the energy $E$ by considering the second identity above and taking $\OO = x^{m-1}$:
\begin{equation*}
	E\langle x^{m-1}\rangle=\frac{1}{2}\langle x^{m-1} p^{2}\rangle+\langle x^{m-1} V(x)\rangle
\end{equation*}
By combining these relations we can eliminate the expectations values of mixed operators $\langle x^np^m \rangle$. The result is a recursion relation for the moments $\ex{n}$ which depends on the energy $E$ of some eigenstate. This recursion relation captures the dynamics of the Hamiltonian:
\begin{equation}\label{eq:recursionfull}
	0=2 m E\langle x^{m-1}\rangle+\frac{1}{4} m(m-1)(m-2)\langle x^{m-3}\rangle-\langle x^{m} V^{\prime}(x)\rangle-2 m\langle x^{m-1} V(x)\rangle
\end{equation}
We note that the $m = 1$ case of this relation is nothing but the virial theorem:
\begin{equation*}
	E = \frac{1}{2}\langle xV'(x)\rangle + \langle V(x)\rangle
\end{equation*}
To use any recursion relation, we need a minimal set $S = \{E,\ex{},\ldots\}$ which can initialize the recursion. We call such a set the `search space'. It will contain the energy and a few moments $\ex{n}$. The dimension $s_* \equiv \dim(S)$ will depend on the potential. For polynomial potentials, the expectation should be that $s_* \sim \deg V(x)/|G|$ where $G$ is any discrete symmetry group of the Hamiltonian (generically $\mathbb{Z}_2)$. Some examples are given below:
\begin{itemize}
	\item $V(x) = \frac{1}{2}\omega^2x^2;\quad S = \{E\}$
	\item $V(x) = gx^3; \quad S = \{E,\ex{},\ex{2}\}$
	\item $V(x) = gx^2 + hx^4;\quad S = \{E,\ex{2}\}$
\end{itemize}
One can show that $s_* = 1$ only for the Coulomb and harmonic oscillator potentials\footnote{A necessary condition is that the virial theorem relates $E = \alpha\ex{n}$ for some $n$, so that determining $E$ determines a moment directly. Secondly, this must be enough data to generate all the other moments.
This only occurs for these two potentials, possibly with an angular momentum barrier if the variable $x$ is properly restricted.}. We always normalize states by demanding that $\ex{0} = 1$. 

\subsection{Positivity constraints and moment problems}
Given a Hamiltonian $H$, we can generate a recursion relation and identify the minimal search space $S$. Then, choosing a point $s \in S$, we can construct a moment sequence from the recursion relation (\ref{eq:recursionfull}). We now need some way of physically accepting or rejecting such a (finite) moment sequence $\{\ex{n}\}_0^N$. We first motivate a construction as in \cite{hart} or \cite{dirac}, relying on the positivity of the norm. Let $\OO = \sum_n c_n x^n$. Then, for any $c_n \in \mathbb{C}$, we have
\begin{equation}
	0 \leq \langle \OO^\dagger \OO \rangle  = \sum_{ij}c^*_i \langle x^{i+j}\rangle c_j \equiv \sum_{ij}c^*_i M_{ij} c_j
\end{equation}
since the quantity is a norm of some state in the Hilbert space. In the above we have defined $M_{ij} = \langle x^{i+j} \rangle$. Considering these as matrix elements of some symmetric matrix $M$, the above constraint can be rephrased as $M \succeq 0$; i.e. $M$ is positive semi-definite. \newline

A matrix $M$ constructed as above is known as a Hankel matrix. Given any real sequence $\{a_n\}_0^\infty$ we may construct a symmetric $K \times K$ Hankel matrix  by $(M)_{ij} = a_{i+j},\ 1 \leq i,j \leq K$. As noted by Lin, Hankel matrices and their connection to positive measures were studied classically by Hamburger and Stieltjes \cite{curto} among others.
A much more detailed account of the problem can be found in the book \cite{schmudgen2017moment}.
 The classical Hamburger moment problem asks: given a sequence $\{a_n\}_1^\infty$, does there exist a positive measure $\mu$ on $\mathbb{R}$ such that $a_n = \int x^n d\mu(x)$?\newline

Hamburger and Stieltjes showed that a necessary and sufficient condition is that the Hankel matrix constructed from the sequence is positive semi-definite for all ranks $K$. 
The key in the proof is showing suffiency of this condition. It was first shown on the half line $\mathbb{R}_+$ by Stieltjes, and later generalized to $\mathbb{R}$ by Hamburger. A further refinement due to by Curto and Fialkow \cite{curto} came much later. They considered the \textit{truncated} problem: given finitely many elements of an ostensible moment sequence $\{a_n\}_1^K$, what are the conditions such that there exists an associated positive measure? The answer is, reassuringly, essentially the same: the Hankel matrix constructed from the sequence must be positive semi-definite up to rank $K$. In fact, the case in which the Hankel matrix is singular is pathological; we will in general use the slightly stronger constraint of positive definiteness to enforce the validity of a moment sequence. The precise conditions for existence of a unique measure can be found in \cite{schmudgen2017moment}. These are usually satisfied in quantum mechanics, as we expect that the wave function decays at least exponentially at large distances for bound state problems. This guarantees the bounds on the moments that produce a proper measure.

\subsection{Algorithmic structure}
We are now prepared to discuss the general structure of our bootstrap. We begin with a Hamiltonian $H$ with some potential $V(x)$, and from that we use (\ref{eq:recursionfull}) to write down a recursion for the moments $\ex{n}$. We identify a minimal search space $S$ and choose a large set of trial points in $X_0 \subset S$. The algorithm follows the steps below. 
\begin{enumerate}
	\item For each point $p = (E,\ex{1},\ldots) \in X_0$, generate $2K-2$ terms of the moment sequence $\{\ex{n}\}$ using the recursion relation. 
	\item From these $2K-2$ terms construct a $K \times K$ Hankel matrix: $\left(M_K\right)_{ij} = \ex{i+j}, 0\leq i,j \leq K-1$ corresponding to each point $p \in X_0$. 
	\item Check positive definiteness of each Hankel matrix. If $M_K(p)\succ 0$, then $p \in X_K$, the set of allowed values at `depth' $K$. If not, throw out the point. 
	\item Obtain a set of allowed values at depth $K$: $X_K \subseteq X_0 \subset S$. Iterate this procedure for larger values of $K$, noting that $X_{K+1} \subseteq X_K$. 
\end{enumerate}
We call the size of the Hankel matrix $K$ the `depth' of the constraint. As $K \to \infty$ we expect the set of allowed points $X_K \subset S$ to converge to points associated with the exact spectrum of the Hamiltonian. 

There are multiple benefits to this general structure. First of all, convergence is monotonic in $K$, in the sense that $X_{K+n}\subset X_K$ for any $n>0$. This follows from the fact that positive definiteness of a Hankel matrix requires positive definiteness of any of its principal submatrices. From a computational perspective this greatly reduces the size of the trial space at each successive iteration. In addition, the steps above are of polynomial complexity in the depth $K$ but, as we shall see, the algorithm displays exponential convergence in $K$. The combination of these factors makes the approach both computationally straightforward and numerically precise.  

There are some issues to be dealt with. The moment sequences of confining potentials tend to grow extremely quickly. In the hydrogen atom, for example, the radial moments $\langle r^k \rangle$ grow factorially. This results in extremely large matrix entries at relatively low depths $K$. Since positive definitness is quite sensitive to the large elements of the matrix, insufficient numerical precision can create errors. However, this same sensitivity to initial conditions is likely what makes the algorithm converge so quickly: a small perturbation $p\in S \to p + \delta \in S$ is liable to greatly perturb the resulting moment sequence at high depths. 

We will comment on specific implementation details in our presentation of examples. As it stands, the algorithm checks an extremely large number of points, as above, by brute force. This requires constructing matrices and evaluating their determinants or performing a Cholesky/LU decomposition at each point. Instead of using this method to find allowed regions of positivity, one could instead look for singular Hankel matrices, which correspond to points lying on the boundaries of the allowed regions. This could, in principle, result in a large computational speedup. We simply mention this approach as a curiosity, first noted by Lin \cite{lin}, that we did not fully explore.  

\section{The Coulomb potential}
Having introduced the generalities, we move on to our first example: the hydrogen atom. Our (radial) Hamiltonian is
\begin{equation}
    H = \frac{1}{2}p_r^2 + \frac{\ell(\ell+1)}{2r^2}- \frac{1}{r}
\end{equation}
Here $p_r$ obeys $[p_r,r] = -i$. We are working with unit mass and $\hbar = 1$. Most data shown will be for orbitals with $\ell > 0$; the s-waves will be discussed independently. From this Hamiltonian we obtain a recursion relation
\begin{equation} \label{eq:hydrec}
    0 = 8mE\langle r^{m-1} \rangle + (m-1)[m(m-2) - 4\ell(\ell+1)]\langle r^{m-3} \rangle + 4(2m-1)\langle r^{m-2} \rangle
\end{equation}
The $m = 1$ case is the virial theorem, which gives the $\er{-1}$ moment in terms of the energy:
\begin{equation}
	E = -\frac{e^2}{2}\left\langle \frac{1}{r} \right\rangle
\end{equation}
As a result this recursion relation closes given just the energy $E$; the search space is one dimensional. From a given energy $E$ we can generate a moment sequence $\{\langle r^k\rangle\}_0^N$ for any $N>0$.

\subsection{Results}
\begin{figure}
    \centering
    \includegraphics[width = \textwidth]{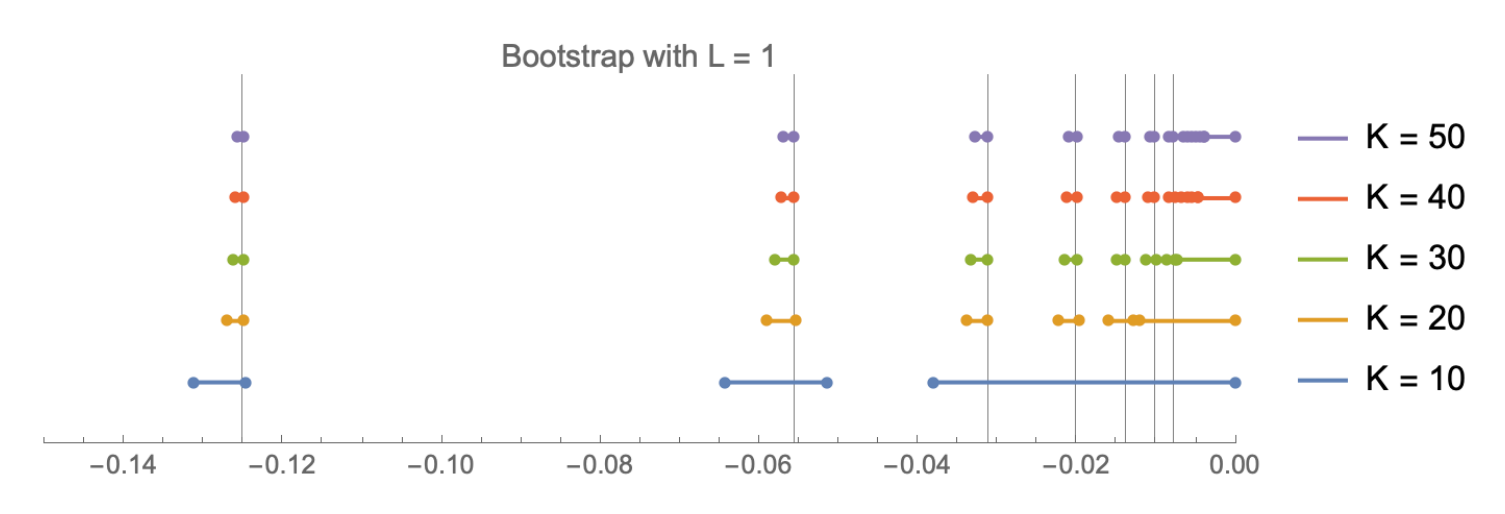}
    \caption{Bootstrap for hydrogen at various $K$ and $\ell = 1$, exact levels in gray. Energy axis pictured.}
    \label{fig:hbtspexample}
\end{figure}

Fig. \ref{fig:hbtspexample} shows what the regions of validity within the search space look like in practice. The axis is the energy $E$, and the intervals pictured are the regions of allowed values at various depths $K$. The convergence as $K$ grows is apparent. Estimates for the allowed energies can be extracted by simply taking the mean of each region at some high depth $K$. The size of the region in question gives a natural notion of error. We go on to show the bootstrapped spectrum and characterize the convergence of the various intervals as a function of $K$. 
\subsubsection{Implementation details}
The results for the spectrum and for the convergence properties were generated by a bootstrap program running with the following parameters in \textit{Mathematica}:
\begin{itemize}
    \item Minimum and maximum depths $K_{min} = 10,\ K_{max} = 50$
    \item Quantum number $\ell$ ranging from $\ell_{min} = 1$ to $\ell_{max} = 10$
    \item Initial step size of $1.499\cdot 10^{-6}$
    \item Initial search space: energies $ E \in [-0.15,-0.0001]$
    \item Numerical precision of 50 digits
\end{itemize}
High numerical precision is required for high depths, as mentioned earlier. At large $m$, the recursion relation (\ref{eq:hydrec}) takes the form
$$
8E\langle r^{m-1} \rangle \approx  (m-1)(m-2) \langle r^{m-3} \rangle + ...
$$
implying an approximately factorial growth. This is consistent with the actual radial moments $\langle r^n \rangle$ of the solved hydrogen model, which are linear combinations of gamma functions in $n$. As a result, even at moderate depth $K$ the matrix elements of $M_K$ can be extremely large in magnitude. To dampen this effect, one can rescale the matrix elements as
$$
M_{ij} \to \frac{M_{ij}}{M_{i1}M_{j1}}
$$
This rescaling preserves the signs of the eigenvalues of the minors, and hence positive definiteness of $M$. Despite this, high numerical precision is required. At larger depths $K$, \textit{Mathematica} machine precision ($\sim 15$ digits) is insufficient to distinguish allowed and disallowed energies. The uncertainty compounds quickly due to the fast growth of terms in the moment sequence, hence the high manual precision. 

By taking the midpoints of the allowed energy intervals at high depths we can extract a bootstrapped spectrum. 
\begin{figure}[h!]
    \centering
    \includegraphics[scale = 0.35]{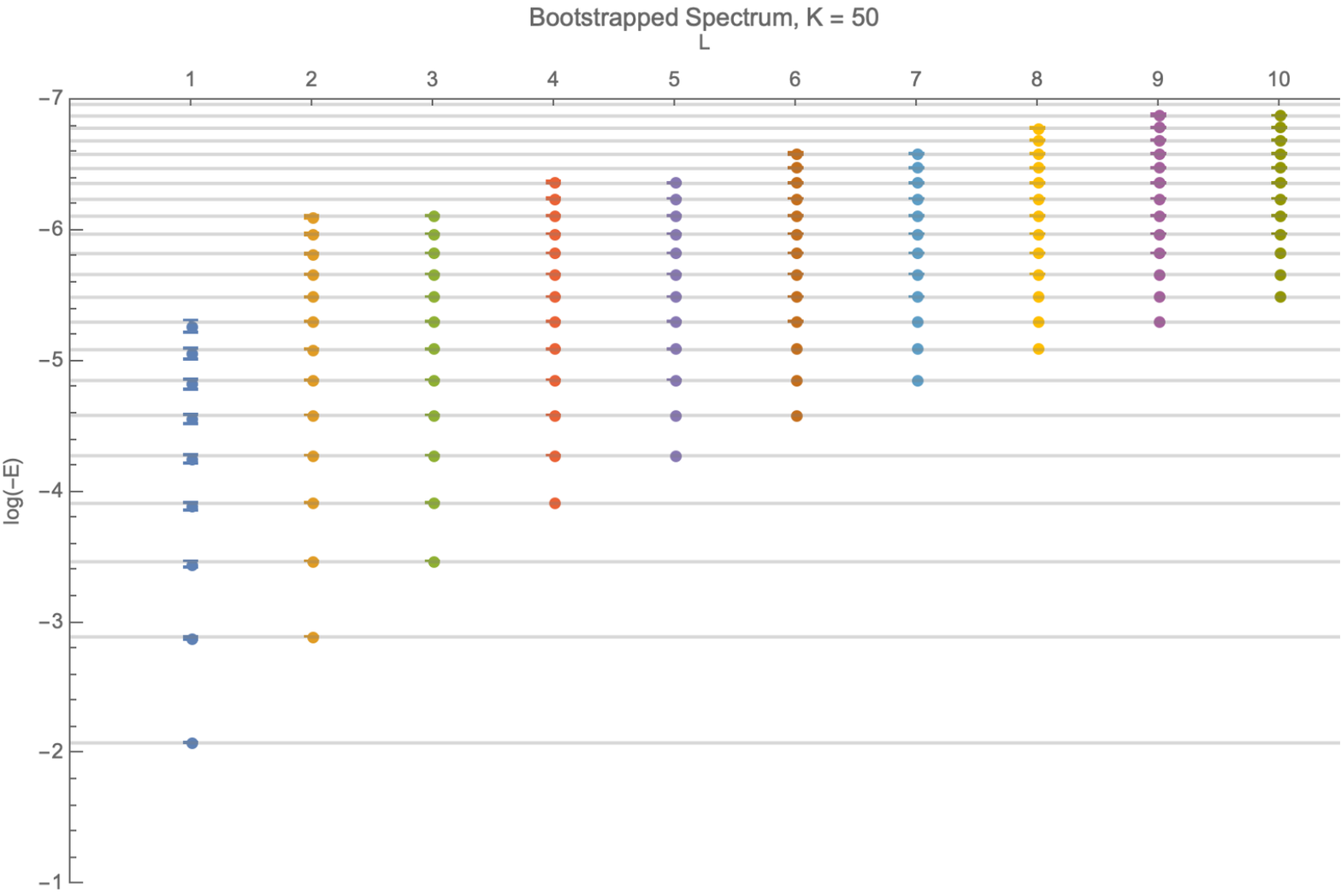}
    \caption{Bootstrapped hydrogen spectrum for various $\ell$, at $K = 50$. Exact spectrum in gray. The energy is displayed on a log scale, so that the energy levels are better distinguished. Error bars included.}
    \label{fig:chemspec}
\end{figure}
\begin{figure}[h!]
    \centering
        \includegraphics[width=6cm]{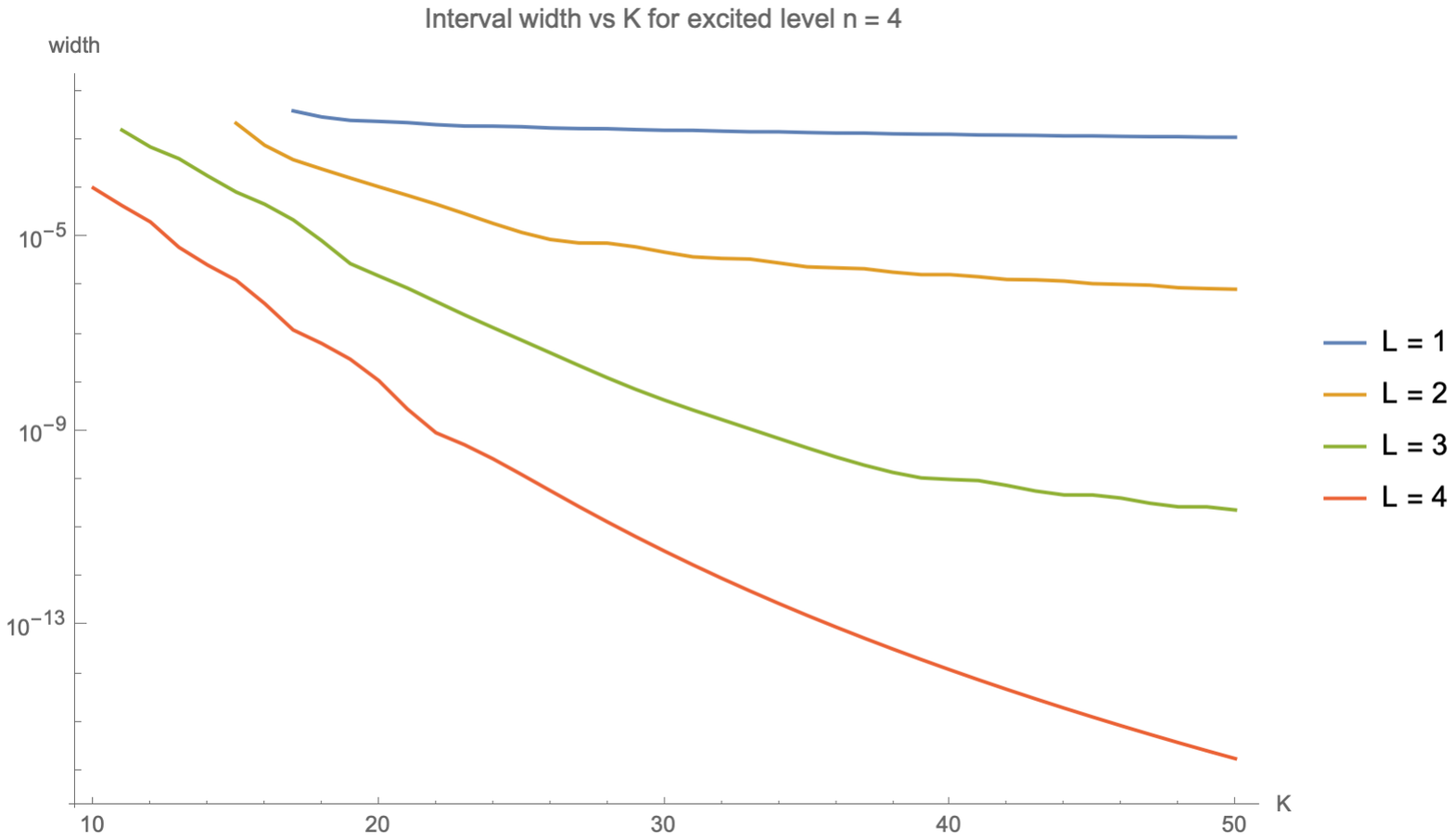}
        \includegraphics[width=6cm]{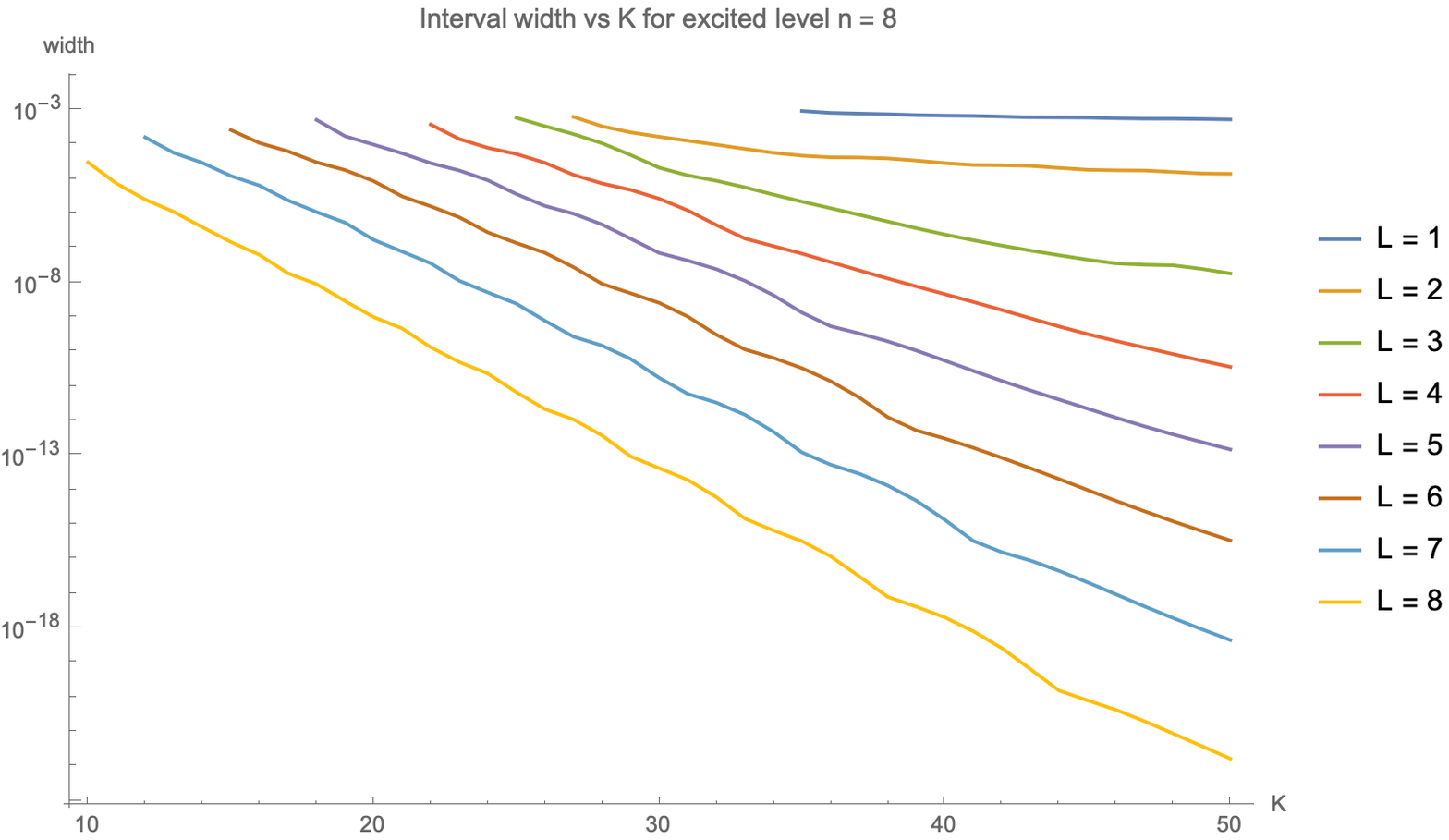}
        \caption{Width convergence of the fourth/eighth excited levels for variable $\ell$. Convergence is very fast for large $\ell$.}\label{fig:conv}
\end{figure}
The width of the intervals decreases exponentially as a function of $K$; this is apparent from Figs. 3, 4 (note the logarithmic scale). Because of this, we zoom in on the allowed values as we iterate further.
 A curiosity is that the exact energies tend to be closer to the more positive side of each allowed interval. In this sense the intervals converge more quickly from the right than from the left. Some examples showing this are relegated to Appendix A. Predictions made by the bootstrap can be extremely precise; a table of bootstrapped predictions with percent errors is also included in the appendix. \newline

While the results from bootstrapping the Coulomb model are encouraging, this is in no way representative of a generic implementation of the bootstrap method. The search space is one-dimensional; this greatly affects the algorithmic structure. But the general strategy remains unchanged, as does the convergence. The real benefit of the method is its simple generalization to matrix models and matrix quantum mechanics, where standard numerical techniques are not as well-known. 

\subsubsection{Issues with $\ell = 0$}

Unlike the $\ell\geq 1$ cases, for $\ell=0$ the naive bootstrap method gives no constraint on the energy, other than $E<0$. Basically, the higher moments all seem to lead to a consistent Hankel matrix.
We do not fully understand the reason for this, but it seems likely that it is caused by the fact that the potential becomes unbounded below at the origin. For $\ell>0$, the origin is missed by the angular momentum repulsion and that seems to stabilize the problem. Notice also in figure \ref{fig:conv}, that the convergence properties for low $\ell$ are much slower than for large $\ell$. This could be a symptom of the failure to converge to energy levels for $\ell=0$.

On the other hand, the hydrogen atom potentials with different values of the angular momentum are related to each other by supersymmetric pairs of Hamiltonians (see for example \cite{Cooper:1994eh}). Consider the pairs of operators
\begin{align}
a\sim&\  i p_r + \frac{A}{r}+B\\
a^\dagger \sim& -ip_r+ \frac{A}{r}+B
\end{align}
It is easy to show that $\frac 12 a^\dagger a$ and $\frac 12 a a^\dagger$ are isospectral, except for the zero modes. Up to a shift, these have the form of the hydrogen atom Hamiltonian, with different values of the angular momentum squared operator $A(A-1), A(A+1)$. Adjusting $A$ and $B$, we can guarantee that we relate two values of the angular momentum.

The important point is that the positivity properties of the moments of $r$ at $\ell=1$, can 
be obtained from acting on wave functions at $\ell=0$ with $a$. That is, we find that the moments 
\begin{equation}
    \langle r^n\rangle_{\ell=1}\sim \langle a^\dagger r^n a \rangle_{\ell=0}
\end{equation}
The operators $a^\dagger r^n a $ is self-adjoint and positive as well. The positivity constraints on $\ell=1$ can be used to solve the $\ell=0$ case, except for the ground state.
That is defined by being a zero mode, which solves $a|\psi\rangle=0$.

\section{The harmonic oscillator}

The recursion relation for the harmonic oscillator is given by
\begin{equation}
    s\langle x^s\rangle = 2 E  (s-1)   \langle x^{s-2}\rangle+\frac{(s-1)(s-2)(s-3)}4 \langle x^{s-4}\rangle
\end{equation}
We use $\langle x^0 \rangle=1$, and $\langle x^2\rangle=E$ to set up the recursion,
while we use $\langle x^{2m-1}\rangle=0 $ for all odd values, 
from the even properties of the potential.

\begin{figure}[ht]
\centering
\includegraphics[width=6cm]{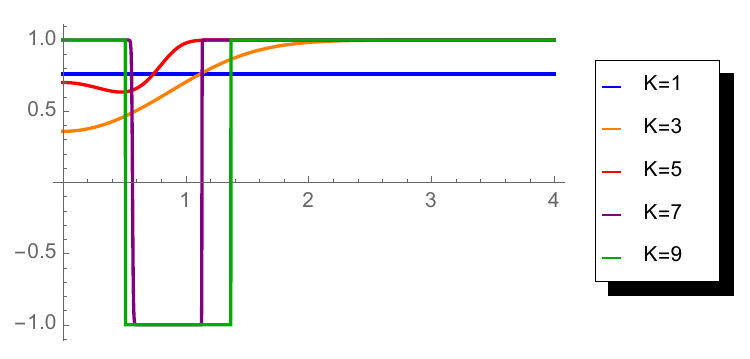}\quad\includegraphics[width=6cm]{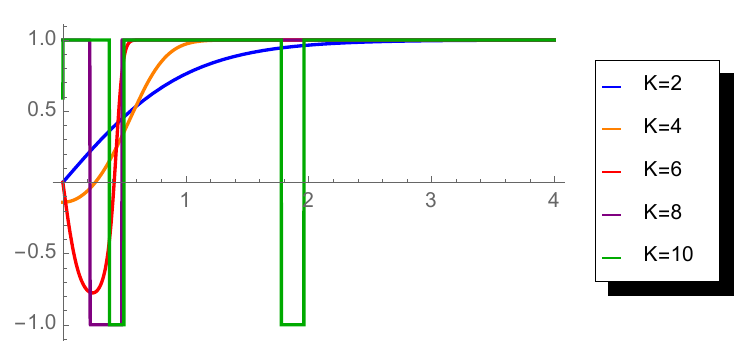}
\caption{Constraint regions for $K\times K$ matrices, split between even columns on the left and odd columns on the right. The horizontal axis is the guess for the energy. We plot $\tanh(\det(M_{e,o}))$ for the different size matrices for added visibility.
Only the region where all curves are positive is valid.
We see that the $K=4,6$ odd matrices are crucial for removing the lower values of $E$ and that negative windows start opening up for larger $K$ that start separating the different eigenvalues.}\label{fig:SHO}
\end{figure}

At first, it might seem that the odd moments of the distribution are superfluous. That turns out not to be the case. The Hankel matrix with the odd moments leads to  additional constraints.
The simplest such constraint is that all even moments are positive. Here we see the example of the $4\times 4$ matrix. 
\begin{equation}
M_{4\times4}=\left(
\begin{array}{cccc}
 1.000 & 0 & E & 0 \\
 0 & E & 0 & 1.500 E^2+0.3750 \\
 E & 0 & 1.500 E^2+0.3750 & 0 \\
 0 & 1.500 E^2+0.3750 & 0 & 2.500 E^3+3.125 E\\
\end{array}
\right)
\end{equation}
We see that there are two independent matrices, one made from the intersection of the even columns and rows, and another from the intersection of the odd columns and odd rows.
The first constraint, from the odd-odd $1\times 1$ matrix $M_{11}$ is that $E\geq 0$. There is no additional constraint from the $2 \times 2$ even matrix made of $M_{00},M_{0,2},M_{2,2}$. 
The results of the constraints are shown in figure \ref{fig:SHO}.

In the figures, we see that the allowed region for energies starts fragmenting as we increase $K$. The determinant functions are determinants of polynomials, so there is a finite number of zeros. It necessarily follows that to see more energies individually, we need to go to higher order in $K$. 

Similar to the Coulomb problem, the convergence is exponentially fast. There are no surprises. Again, we need high precision because the moments grow factorially.
\begin{figure}[ht]
\centering
\includegraphics[width=8cm]{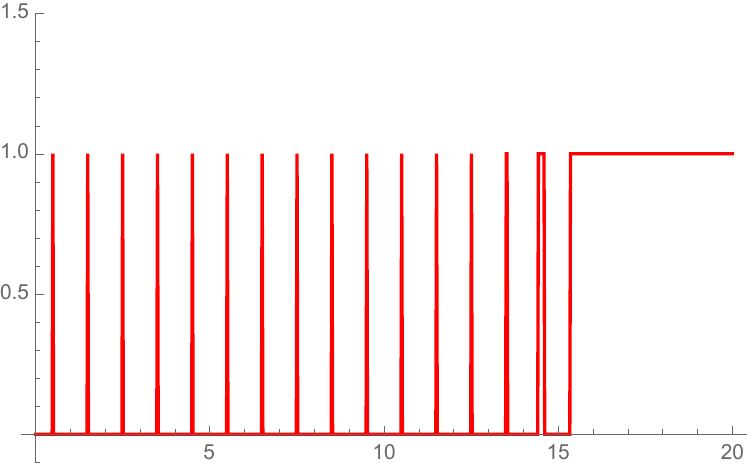}\caption{Allowed Harmonic oscillator energies at level $K=55$. The peaks are small windows at the half integers. The horizontal axis is the energy, and the vertical axis is an indicator function if a point is allowed or not. Fifteen energy levels are resolved. }\label{fig:allowedHosc}
\end{figure}

\section{Conclusion}

In this paper we have investigated the bootstrap method for quantum mechanics against problems that admit an analytical solution and which depend only on one parameter for the bootstrap problem. This is done for the hydrogen atom and the harmonic oscillator. 

The bootstrap is generated by the moments of the wave function at a given energy, which satisfy a recursion relation. The problem one needs to solve is consistency of the Hankel matrix, which needs to be positive definite.

Overall, the method works very well. However, we found one case where the method does not give useful information on its own, for the hydrogen atom when the angular momentum vanishes. This seems to be related to the fact that  the potential is unbounded from from below. We found, using supersymmetric Hamiltonians, that one could in principle solve a different positive matrix problem recursively to get the $\ell=0$ states.

Energy levels are resolved individually, in a sequential manner as  one increases the size of the Hankel matrix. The convergence to the exact answer is very fast. Given the data, it seems like the convergence is exponential. Convergence was also not uniform: different values of angular momentum gave very different convergence rates, with higher angular momentum converging faster. The data seems to converge only to allowed energy levels and we found no exotic solutions to the constraints. It would be interesting to understand 
if there is a theorem that controls this behavior, and under what conditions it does so.

For the problems studied, the moments grow factorially. The check for positivity of the Hankel matrix requires very high precision to verify. 

It is interesting to extend this analysis to other problems. Currently we are looking at the double well potential, where the number of parameters to fix the bootstrap problem increases from one to two. We are also considering models with non-polynomial potentials, where trigonometric moments must be considered. 

One could also investigate how these techniques can be used to understand not just the energy levels and the moments of distributions at fixed energy, but also more general matrix elements between energy levels. We are currently looking into this possibility.

\acknowledgments
We would like to thank R. Brower, S. Catterall, X. Han, Y. Meurice  for discussions. Research supported in part by the Department of Energy under grant DE-SC0019139.

\appendix
\section{Hydrogen: additional results}
Here we include some results from the hydrogen Hamiltonian bootstrap. We commented earlier on the lopsided convergence of the intervals. The following plots track the difference between the left/right ends of the intervals and the exact value for the energy. Despite some oscillation, the right (dashed) converges more quickly than the left (solid).
\begin{figure}[h!]
    \centering
    \begin{minipage}{0.45\textwidth}
        \centering
        \includegraphics[width=\textwidth]{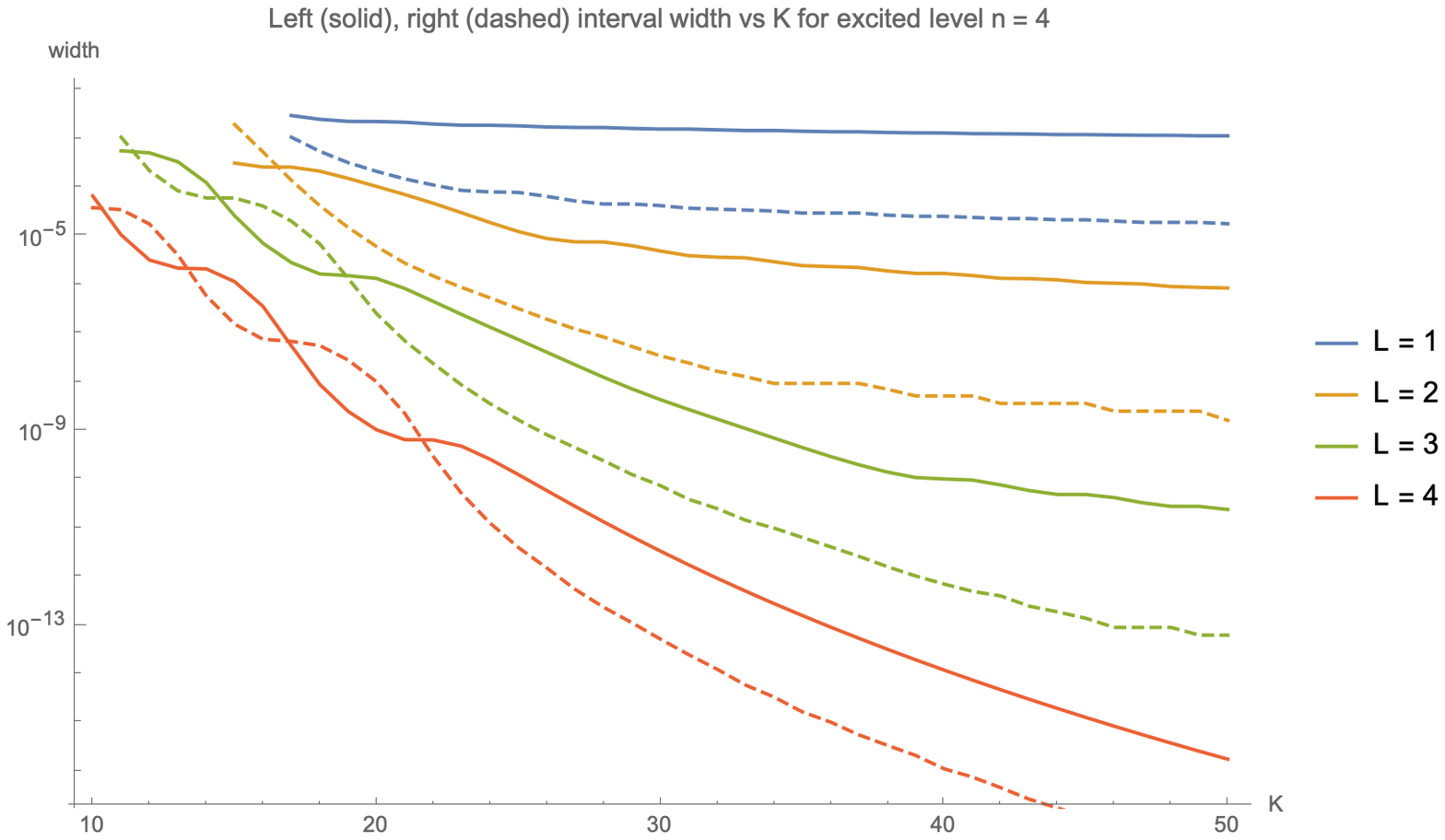}
        \caption{Left/right convergence properties for the fourth excited level.}
    \end{minipage}\hfill
    \begin{minipage}{0.45\textwidth}
        \centering
        \includegraphics[width=\textwidth]{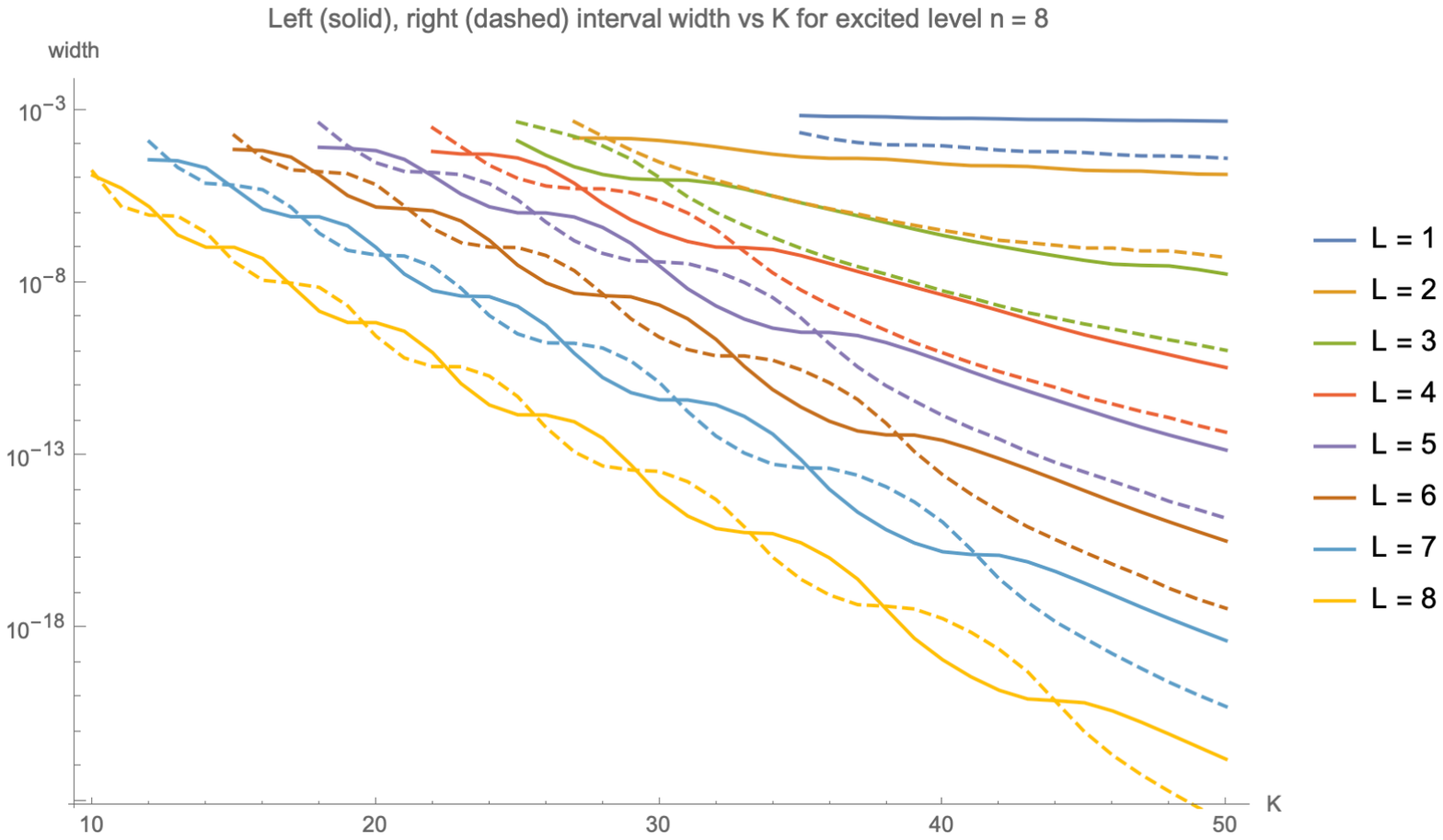}
        \caption{Left/right convergence properties of the eighth excited level.}
    \end{minipage}
\end{figure}
This, and the earlier convergence figures, shows that higher values of $\ell$ converge more quickly. Given this, to get the best estimate for all energy levels, one should choose the lowest level found for each value of $\ell$. Carrying this out (and using the maximum $\ell = 10$ for levels higher than $n = 10$), the bootstrap at depth $K = 50$ detects 22 energy levels, all with below 0.3 percent error. The included table gives the results of this bootstrap, and cites the central, left, and right percent error versus the true value. The left and right percent errors are calculated using the left and right boundaries of the interval found around the correct energy level $n$. \newline

The percent error begins to grow after $n = 10$; this is a reflection of the way the energies were sampled. For $n \leq 10$, the energy is calculated with $\ell = n$, which supplies the best convergence properties for that level. Once $\ell = 10$, the maximum value, the program looks for the higher energy levels detectable at $l = 10$. As these energies approach zero, higher resolution is needed and the relative error at $K = 50$ grows.
\begin{figure}[!h]
    \centering
    \includegraphics[width=\textwidth]{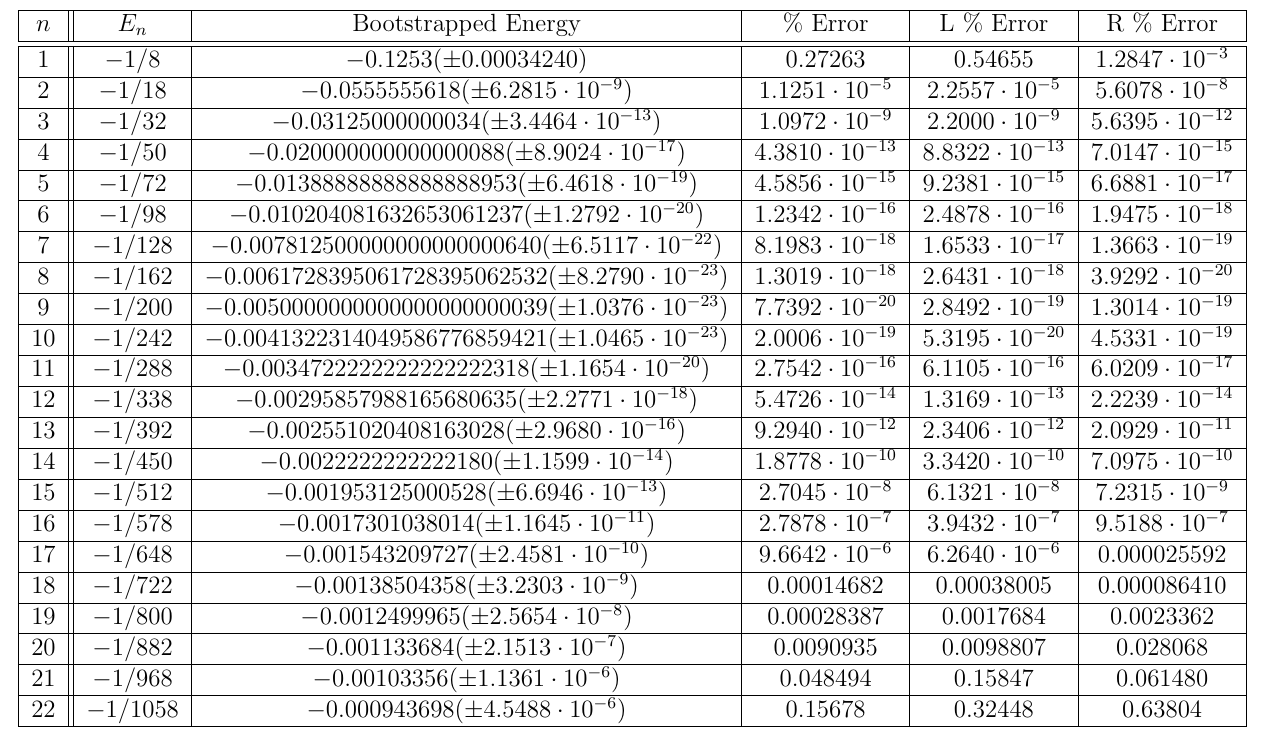}
    \caption{Comparison of exact and bootstrapped energy levels. We use the best result for each value, with a bound on $\ell\leq 10$. The percent error included. L/R percent error compares the left/right sides of the interval to the correct value. The exact value for $E_n=-(2 (n+\ell)^2)^{-1}$ has the principal number $n$ shifted by one since we start at $\ell=1$, rather than $\ell=0$.}
    \label{fig:my_label}
\end{figure}


\bibliography{refs}
\bibliographystyle{JHEP}

\end{document}